\documentclass[twocolumn,amsmath,amssymb,pra,showpacs,superscriptaddress,aps]{revtex4}

\usepackage[english]{babel}
\usepackage[applemac]{inputenc}
\usepackage[T1]{fontenc}
\usepackage{verbatim}
\usepackage[pdftex]{graphicx}
\usepackage[normalem]{ulem}
\usepackage{color,braket}

\definecolor{dgreen}{rgb}{0.0, 0.7, 0.0}
\definecolor{mycolor}{rgb}{0.9, 0, 0.9}

\begin{document}
\bibliographystyle{aip}

\title{Occupation-constrained interband dynamics of a non-hermitian two-band Bose-Hubbard Hamiltonian}

\author{Manuel H. Mu\~noz-Arias}
\affiliation{Departamento de F\'isica, Universidad Del Valle, A. A. 25360, Cali, 
Colombia}
\author{Carlos A. Parra-Murillo}
\affiliation{Departamento de F\'isica, Universidad Del Valle, A. A. 25360, Cali, 
Colombia}
\author{Javier Madro\~nero}
\affiliation{Departamento de F\'isica, Universidad Del Valle, A. A. 25360, Cali, 
Colombia}
\affiliation{Centre for Bioinformatics and Photonics-CIBioFi, Calle 13 No. 100-00, Edificio 320, No. 1069, Cali, 
Colombia}
\author{Sandro Wimberger}
\affiliation{Dipartimento di Scienze Matematiche, Fisiche e Informatiche, Universit\`{a} di Parma, Parco Area delle Scienze 7/a, 43124 Parma, Italy}
\affiliation{INFN, Sezione di Milano Bicocca, Gruppo Collegato di Parma, Parma, Italy}
\affiliation{ITP, Heidelberg University, Philosophenweg 12, 69120 Heidelberg, Germany}

\begin{abstract}
The interband dynamics of a two-band Bose-Hubbard  model is studied with strongly correlated 
bosons forming single-site double occupancies referred to as doublons. Our model for resonant doublon
interband coupling exhibits interesting dynamical features such as quantum Zeno effect, the generation of
states such as a two-band Bell-like state and an upper-band Mott-like state. The evolution of the asymptotic
state is controlled here by the effective opening of one or both of the two bands, which models decay channels.
\end{abstract}

\date{\today} 

\maketitle

\section{Introduction}
\label{sec:introduction}

The Bose-Hubbard (BH) Hamiltonian is a celebrated many-particle model describing 
atoms trapped in a periodic array of potential wells \cite{zoller}. 
The simplest version of this model, the single-band Bose-Hubbard Hamiltonian, already 
accounts for a plethora of interesting physical phenomena which have been investigated 
in experiments with ultracold quantum gases and optical lattices formed by laser 
beam interference patterns \cite{bloch}. These setups provide a high degree of system control 
as, for instance, of the lattice depth that rules the inter-well particle
tunneling, and of interparticle interaction tuned via Feshbach resonances \cite{bloch}. 
The combination of this two components permits the implementation of the remarkable idea 
of quantum simulation that allows for the understanding of the underlying physics of complex systems 
in a clean manner \cite{bloch,greiner}. 
However, a whole set of physical phenomena are left away within the single-band approach. An example is the Landau-Zener transitions 
reported in Refs.~\cite{sias,pisa,zenopisa} and theoretically tackled using mean field methods \cite{meanfield}. 
Yet, the limit of strongly correlated particles is more challenging to treat 
since the Bose-Hubbard Hamiltonian extended to higher bands contains a 
lot of new single and many-particle processes rendering the system overwhelmingly complex \cite{ploetz,parra}.

The effects of higher Bloch bands on the single-particle dynamics \cite{kolovsky}  and 
on the spectral statistics of the single-band approach \cite{tomadin} were studied. 
One may address the problem by restricting the Fock space to a two-band Bose-Hubbard Hamiltonian,  
where the interband coupling is induced via an external field \cite{ploetz,parra}. 
In this paper we are interested in the two-band Bose-Hubbard Hamiltonian 
assuming strong correlation between the particles constraining the single-site 
interband coupling to occur only when there is more than one particle in one site. 
Our constrained model, that allows only for specific interband tunneling processes, is a special situation
of the full two-band model recently studied in the different context of periodic driving \cite{parra16}.

We explore the dynamical possibilities arising from coupling the system to 
decay channels, e.g. corresponding to higher energy bands, which are neglected in our model.
Thus, the effective "reservoir" consists of the continuum part of the Wannier-Stark spectrum \cite{kolovsky,sias,pisa}. 
Our motivation to explore the dynamics of an open two-band system is inspired by 
recent works reporting emergent phenomena induced by loss. 
Examples are the work on quasi-dissipation free subspaces of 
entangled atomic Bell-like states \cite{parra16}, 
localization of BECs via boundary dissipation \cite{oppo}, 
entanglement detection~\cite{kollath}, the possibility 
of controlling many-body quantum dynamics by localized dissipation \cite{ott1,witthaut}. 
Dissipation replaces the continuous observations required 
to observe a quantum Zeno effect  \cite{misra,zeno,ott2,zenopisa,newzeno}. 
The engineering of controlled onsite dissipation channels has been developed in recent 
experiments \cite{wurtz,ott2} such that the range of possible experimental applications is continuously increasing.
Modern experimental methods may as well allow for addressing individual bands in a lattice by state-selective excitation processes, see e.g.
\cite{sias}. The motivates the application of decay channels with independent rates in our system.

This work is organized as follows. In Sec. \ref{sec:physical_system}, we introduce 
our constrained two-band BH model to study correlated interband dynamics. We also sketch 
how our model can be extracted from the full two-band BH model \cite{parra} and the specific set of 
interband phenomena it aims to represent. In Sec. \ref{sec:numerical_results}, we briefly present the methodology  
to study correlated interband dynamics. Sec. \ref{ssec:zeno} presents results regarding the emergence of the quantum Zeno 
effect in the interband transport, and the conditions under which it appears. In Sec. \ref{ssec:state_preparation}, 
we show how by controlling the ratio between the dissipation rates of the bands one can prepare 
interesting superposition states, in particular those Bell-like states 
already found in \cite{parra16}. Finally in Sec. \ref{sec:conclusions} we present a brief summary and conclusions of our work.

\section{Physical system and Hamiltonian}
\label{sec:physical_system}

\begin{figure}[ht]
\includegraphics[width=\columnwidth]{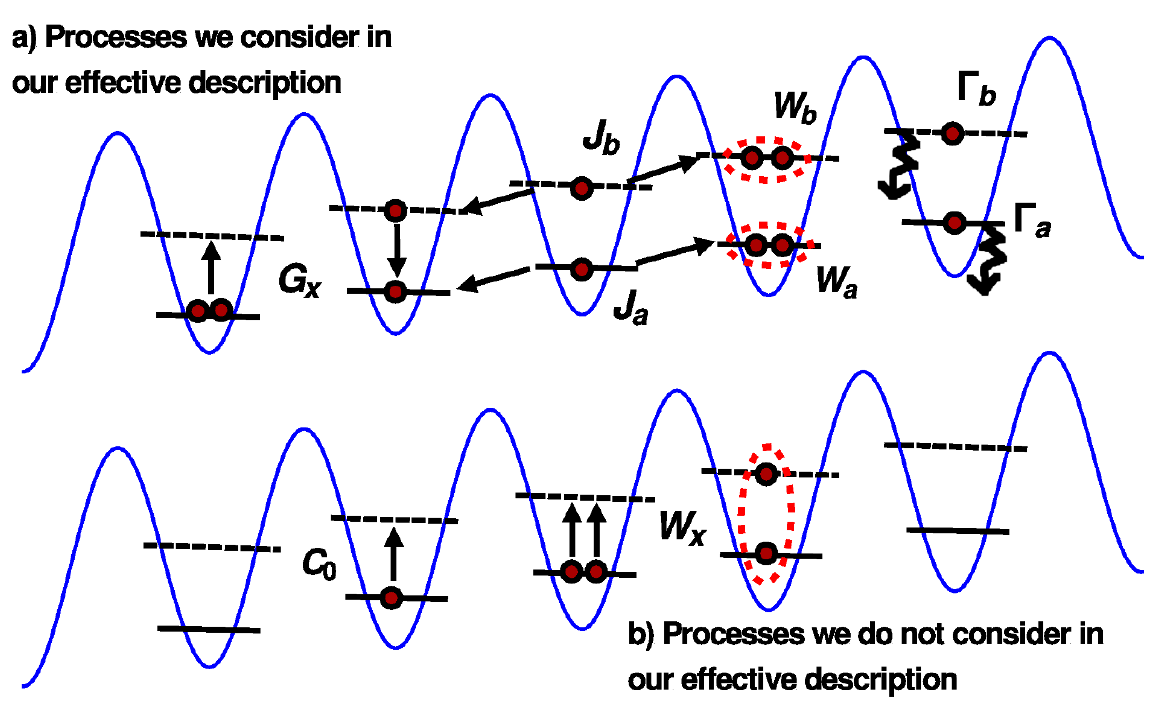}
\caption{\label{fig:1} (Color online) (a) Representation of the different processes 
that take place under the Hamiltonian in Eq.~(\ref{eqn:01}). Particles can jump 
between next neighboring sites with an amplitude $J_{a,b}$, doubly or higher 
occupation of the lattice sites have a increment in energy proportional to 
$W_{a,b}$, the external field contributes with an increment in the energy 
proportional to the magnitude of the external field ($F$) and the position the 
lattice. The inter-band transport occurs with an amplitude $G_x$, and 
$\Gamma_{a,b}$ stands for the decay rates of the two Bloch bands. 
(b) Usual complementary terms of the full two-band BH model 
\cite{parra} that we are not taking into account within our 
effective description. Notice that, under the regime of parameters we are working, these 
processes are well represented by the $G_{a(b)}$ term. Below we show, for instance, that including 
the $W_x$ term almost does not affect the dynamics.}
\end{figure}

Our Hamiltonian, see Eq. \eqref{eqn:01} below, is a restricted model of a much more complex system introduced and 
discussed, e.g., in our previous works \cite{parra,parra16}. It has the great advantage that the complexity is much 
reduced taking into account only the dominant interband couplings under specific resonant tunneling conditions.
The first part corresponds to two independent copies of the tilted single-band BH chain, 
with the external field with magnitude $F$ defining the Wannier-Stark force. 
The second part describes interband coupling restricted to single-site occupation number, i.e., 
there will be transitions only if an individual site is occupied by more than one particle.
The dynamics is induced by two transport processes: first, the intraband hopping of 
particles between next-neighboring sites, and second interband tunneling of so-called 
doublons \cite{Andrey,sachdev,kolovsky2}. Altogether, these processes are described by the following effective Hamiltonian:
\begin{eqnarray}\label{eqn:01}
\hat{H} &=& 
\sum_{l,\,\alpha=\{a,b\}}\left[\varepsilon_l^{\alpha}\hat{n}_l^{\alpha}-\frac{J_{\alpha}}{2}(\hat{\alpha}_{l+1}^\dagger 
\hat{\alpha}_{l} +{\rm  H.c.})+ \frac{W_\alpha}{2} 
\hat \alpha_l^{\dagger\,2}\hat\alpha_l^2 
 \right] \notag\\
&+& \sum_l \left[ G_a\,\,\hat{b}_l^\dagger \hat{a}_l \hat{n}_l^a(\hat{n}_l^a-1) +
  G_b\,\,\hat{a}_l^\dagger \hat{b}_l \hat{n}_l^b(\hat{n}_l^b-1)+
  {\rm  H.c.}\right]\notag\\
&+&\sum_l W_x\, \hat n^a_l\hat n^b_l
\end{eqnarray}
with $$\varepsilon_l^\alpha=\left[d_L F l +\frac{\Delta_g}{2}(\delta_{\alpha,b}-\delta_{\alpha,a})-i \Gamma_\alpha\right] \hat{n}_l^\alpha\,. $$
Here, $\alpha=a\,(b)$ refers to the lower (upper) band,  $\hat{\alpha}_l^\dagger (\hat{\alpha}_l)$ are bosonic creation (annihilation) 
operators with $J_{\alpha}$ the hopping amplitude of each band, and $\hat{n}_l^{\alpha}$ 
the number operator. The on-site interparticle interaction strengths are $W_{\alpha}$  
and $\Delta_g$ stands for the on-site energy bandgap. The interband coupling is represented
  by the $G$-terms. Such a coupling implies that only when there are more than one particle in a single site
  anywhere in the lattice one particle will be promoted to the other band. Its respective counterpart
  implies the increase of a single-site occupancy wherever there is at least one particle
  (see $G_x$ process in Fig.~\ref{fig:1}-(a)). In the case of unitary filling condition for
  the lattice, i.e., $N/L=1$ for any $N$, the most probable higher occupation of a single lattice site is a double
  occupancy, therefore, it is expected to get doublon-mediated interband dynamics. In addition, we have an external
  Stark field $F$ that helps us to tune different resonant regimes \cite{parra16}.

In Fig.~\ref{fig:1}-(a) we present a scheme of the different processes considered
in the effective model, and of those that we do not take into account from the original
full Hamiltonian (see Fig.~\ref{fig:1}-(b)). Note that the Hamiltonian $\hat H$
is not hermitian since the unperturbed energies contains the imaginary number 
$-i\Gamma_{\alpha}$ representing the effective opening of the system, i.e., decay
rates for every band. This implies that the eigensystem of $\hat H$ consists
of a set eigenvalues of the form $\varepsilon_i=E_i-i\Gamma_i/2$ and metastable 
eigenstates with the lifetime $\tau_i\sim 1/\Gamma_i$.

A resonant tunneling between two- and single-particle levels is expected as a consequence of the interband 
$G_x$ coupling term in Eq.~(\ref{eqn:01}). Here, there is no need for a time-dependent 
driving as in Ref. \cite{parra16} and $G_x$ is taken constant in what follows. 
We consider the following possibilities for the initial conditions for our time evolution:
\begin{eqnarray}
(i) & F_0 = W_a/2\pi,& |\psi_0\rangle = |111\cdots;000\cdots\rangle,\notag\\
(ii)& F_0= 0, & |\psi_0\rangle = |2020\cdots,000\cdots\rangle.\notag
\end{eqnarray}

The difference between the initial states is just the presence of the Stark 
field at $t=0$. The two initial states can be dynamically connected by the 
hopping term \cite{sachdev,kolovsky2}, hence the time evolution is expected to be 
equivalent, if not the same. We consider decay rates 
$\Gamma_\alpha \ll J_\alpha$, where $J_\alpha\ll \{U_\alpha, \Delta_g\}$. This means that decay is slow  
and the total particle number tends to be preserved with high probability. 
Then our effective Hamiltonian approach is a good approximation to more complex master equation approaches, see e.g. \cite{kordas,daley}.

In all numerics we compute the time evolution operator $\hat{U}_t = \hat{\mathcal T}\exp\left[-i\int_0^t\hat H(t')dt'\right]$. 
With the help of  $\hat{U}_t$, we compute the time-evolved many-body state starting from a given initial 
state $|\psi_0\rangle$. The figure of merit to study the interband
dynamics is the population inversion, i.e., the population imbalance between the bands defined by
\begin{equation}
 M_t = \langle \psi_t|\frac{1}{N} \sum_l (\hat{n}_l^b-\hat{n}_l^a)|\psi_t\rangle/\langle \psi_t|\psi_t\rangle \, .  
\end{equation}

Our study is divided in two parts. We first consider the situation in which
only the upper Bloch band is open, i.e., $\Gamma_a = 0$ and $\Gamma_b \ne 0$.
In this limit, it will be shown that quantum Zeno-like dynamics occurs 
induced by the dissipative process during the promotion of particles to the
upper band. Secondly, we explore the dynamical consequences of simultaneously
opening both Bloch bands what results in the formation of a Bell-like state.

\section{Numerical Results}
\label{sec:numerical_results}

\subsection{Quantum Zeno effect in the inter-band dynamics}
\label{ssec:zeno}

A quantum system prepared in a meta-stable state can be dynamically locked/frozen 
by means of continuous observations \cite{misra,ott2} or by the control of the system 
via temperature and entropy \cite{erz}.  This phenomenon, known 
as quantum Zeno effect, is a paradigm of the measurement process in 
quantum mechanics. We now show how the quantum Zeno effect occurs in our system 
thus changing radically the expected interband transport (See Refs.~\cite{ploetz}). 
\begin{figure}[ht]
\includegraphics[width=\columnwidth]{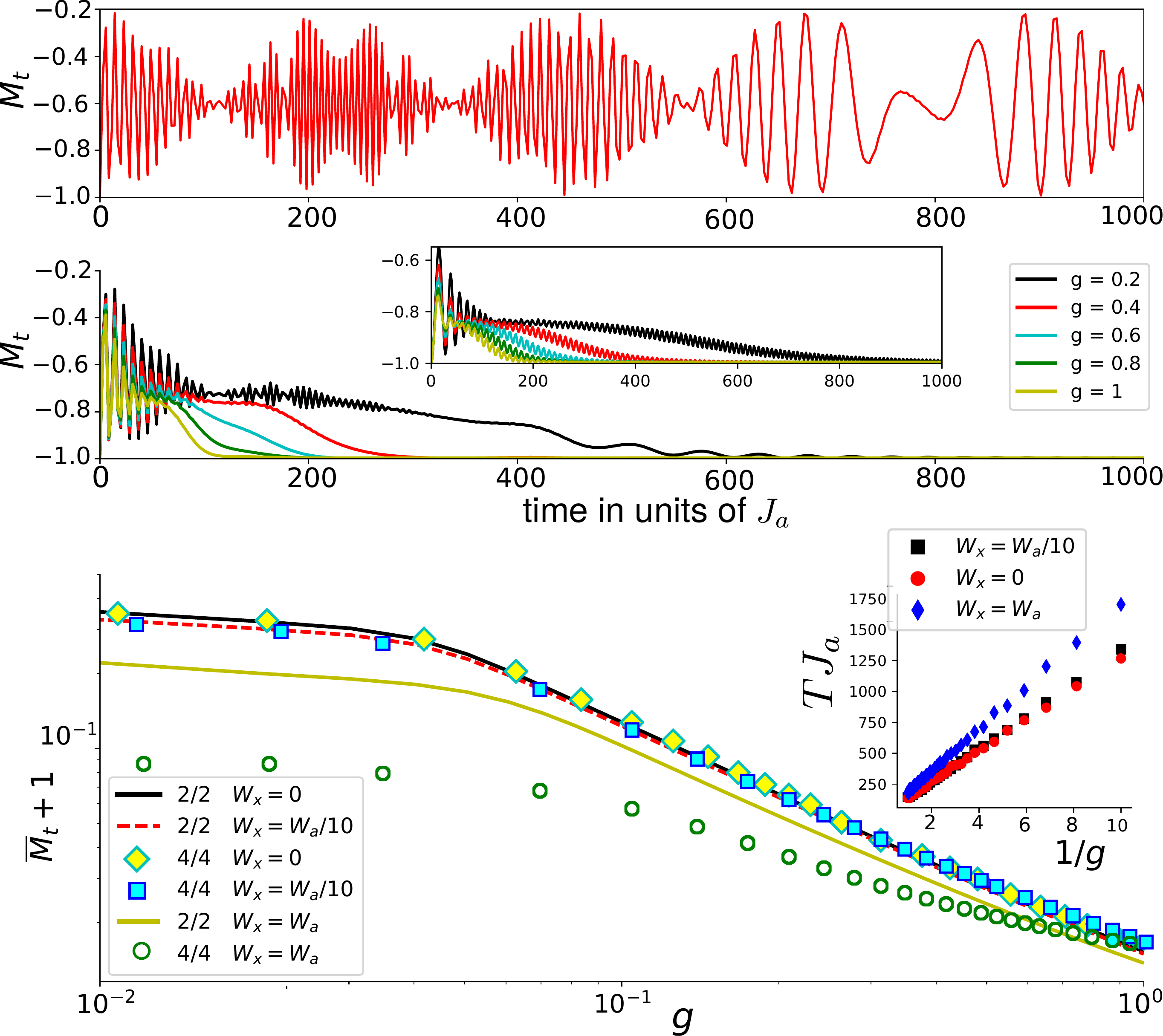}
\caption{\label{fig:2} (Color Online) (a) Time evolution of the population inversion $M_t$ for 
a closed system $\Gamma_a = \Gamma_b = 0$. (b) $M_t$ for different 
values of $\Gamma_b$, we have taken $\Gamma_b = g \times 0.8\times 10^{-3}$, with $g$ a control 
parameter. We see a fast decay of $M_t$ as we increase $g$, in all the cases 
the system ends in the state $|11;00\rangle$. (c) Long time average of 
$M_t$ as a function of $g$. We have plotted this curve in $\log$-$\log$ 
scale after shifting $\overline{M}_t$ by one. The 
red-continuous line is for the $N/L=2/2$ system and the black-dotted line for the 
$N/L=4/4$ system. The tale of both curves follows a power law, evidence of the 
quantum Zeno effect. The inset presents the decay time $T$ of $M_t$ as 
function of $1/g$, the linear behavior is evident. In addition we include the 
term containing the interparticle interaction between particles in different 
bands with intensity $W_x$ to show that the results are practically the same,
see for instance: red-dashed line, light-blue squares. In all the 
computations we use the BH parameters: 
$J_a = -J_b = 0.006$, $W_a = W_b = 0.034$, $W_x\approx (0.1 \ldots 1)\times W_a$, 
$\Delta_g = 0.108$, $G_a = 0.09$ and $G_b=0$}.
\end{figure}
\begin{figure}[!]
\includegraphics[width=0.95\columnwidth]{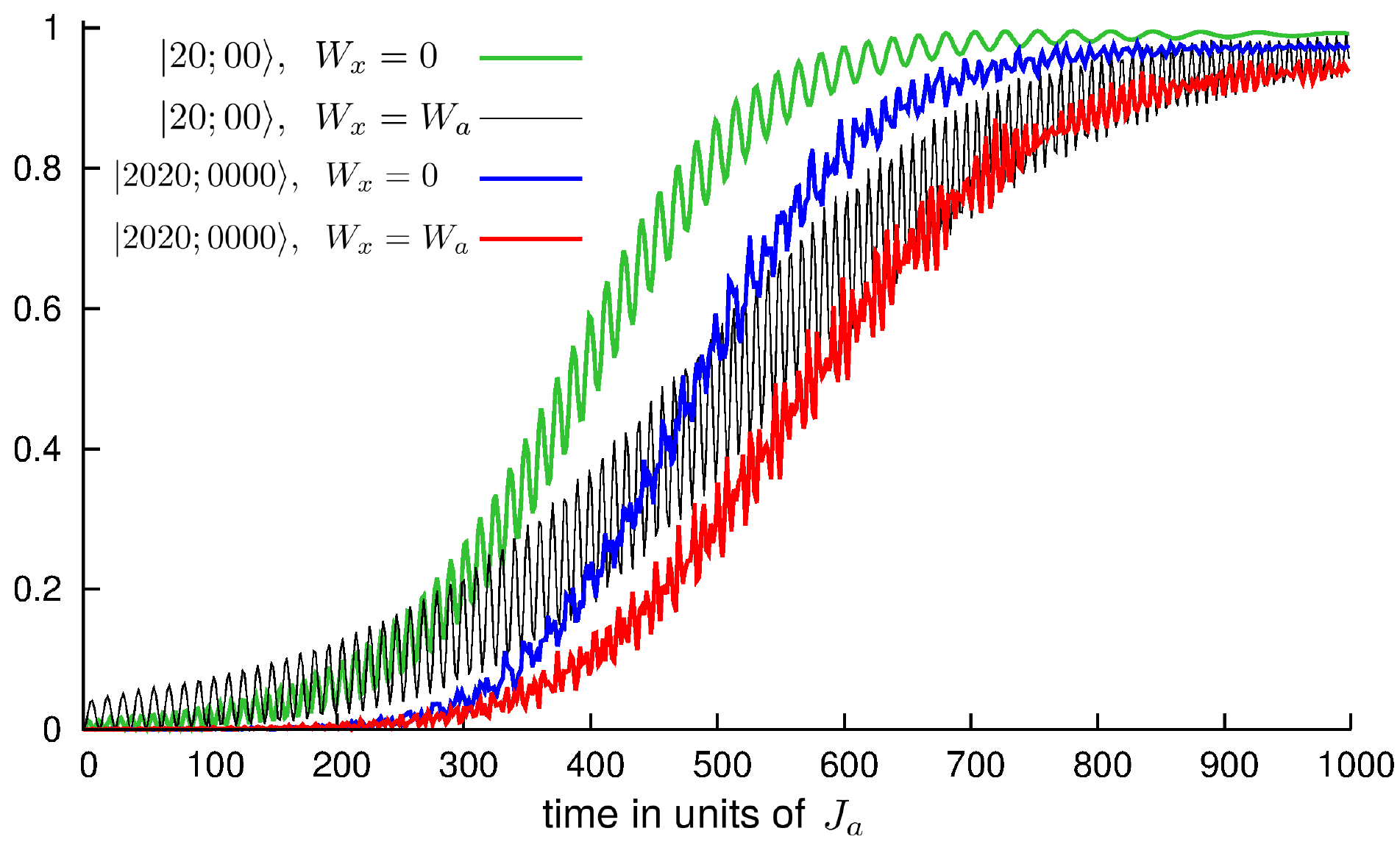}
\caption{\label{fig:2a}(Color Online) {\em Assymptotic state}: The figure 
shows the projection probability on the lower-band Mott state $\ket{111\cdots;000\cdots}$ 
in time where it can be seen that it assymptotically grows to one as mentioned 
in the main text. The time evolution is computed for initial states of the type 
(main panel) $\ket{2020\cdots;0000\cdots}$. The parameters are the same as in Fig.\ref{fig:2}
above; the $W_x$ values are given in the legend.}
\end{figure}

Let us start considering the decay rates only in the upperband, i.e., $\Gamma_b \neq 0$ 
and $\Gamma_a = 0$, and the system defined by the filling factor $N/L=\{2/2,4/4\}$ corresponding 
to $N$ particles in $L$ lattice sites. The initial state is taken as the lower-band Mott state 
$|111\cdots;000\cdots\rangle$ which is dynamically evolved with a fixed tilting 
$F_0=W_a/2\pi$ ensuring the resonant activation of the interband dynamics. This occurs 
in a time scale in which many hopping processes are allowed, leading to occasional 
double occupancies (doublons) in the lower band. Let $g$ be the dissipation control 
parameter, i.e. $\Gamma\propto g$. In Fig.~\ref{fig:2}-(b) we see that the interband 
dynamics shows collapses and revivals for $g=0$ at certain times, similar to those predicted in 
Ref.~\cite{ploetz}. Here we have neglected the interband coupling term $G_b$ since it does
not change relevantly the results of this section. Yet, setting $g\neq 0$ we see that the role of the dissipation is 
to freeze the dynamics and the initial state, $\ket{111\cdots;000\cdots}$, is assymptotically
recovered (see Fig.~\ref{fig:2a}). We stress that our Hamiltonian, see the sketch in Fig.~\ref{fig:1}, 
couples both bands, and that the decay occurs from both bands to outside the system. Hence, the observed 
suppression of upper-band population is non trivial. Indeed, it reflects nothing but the onset 
of the quantum Zeno effect \cite{zenopisa,zeno,newzeno,misra}. 

To see this in a clean way, we compute the long-time average of $M_t$ defined as 
$\overline{M}_t=\lim_{\tau\gg 1/|J_a|}\int_0^\tau M_{t'} dt'/\tau$. In Fig.~\ref{fig:2}-(c) 
we plot $\overline{M}_t$ as a function of the dissipation strength $g$ (continuous line). 
The tail of this curve follows a power-law decay $\overline{M}_t\propto g^{-1}$. Note that
this behavior is preserved even in the present of the interband coupling term: 
$W_x\sum_l \hat n^a_l \hat n^b_l$, (see panel Fig.\ref{fig:1}-(b)) where $W_x= (0.1 \ldots 1)\times W_a$.
Additionally, the inset shows the decay time $T$ of the signal $\overline{M}_t$ as a 
function of $1/g$, with a linear growth of $T\propto 1/g$, a clear signature of 
the quantum Zeno effect \cite{kollath, zeno}. The dots in Fig.~\ref{fig:2}-(c) 
correspond to the results for $N/L=4/4$. We can see that the effect is still preserved
for larger system sizes. In Fig.~\ref{fig:2a}, we show how the initial condition, the Mott state 
$\ket{111\cdots;000\cdots}$, is dynamically recovered after some time. For short, we initialize
 the system one step forward, that is, we directly evolve states with double occupancies 
$\ket{2020\cdots;0000\cdots}$. Note that including the term $W_x$ just delays the effect 
rather than washing it out.

\subsection{Asymptotical dynamics: two-band Bell-like state and upper-band Mott state}

\label{ssec:state_preparation}
To constrain the dynamics of a quantum system to a convenient subspace of the full Hilbert space is one of the hardest but, 
nevertheless, interesting open problems in quantum control theory. Besides, dissipation has 
always been related to the onset of decoherence which induces a strong mixing of a 
large number of accessible states. Yet, there have been advances towards quantum control by means 
of controlled dissipation processes, see e.g. \cite{kollath, daley, witthaut}. This has found applications
in the emerging field of atomtronics \cite{seaman,exp-atomtronics}.

    Here, we show how the asymptotic dynamics governed by Eq.~(\ref{eqn:01}). First, we study the case
  $G_b=0$ that yields a steady state with a structure resembling a Bell state. Secondly, we show that including $G_b\neq 0$,
  the asympotic state takes the shape of a Mott state in the upper band. The 
latter result gives a possible way of creating  such states.

\begin{figure}[hb]
\includegraphics[width=\columnwidth]{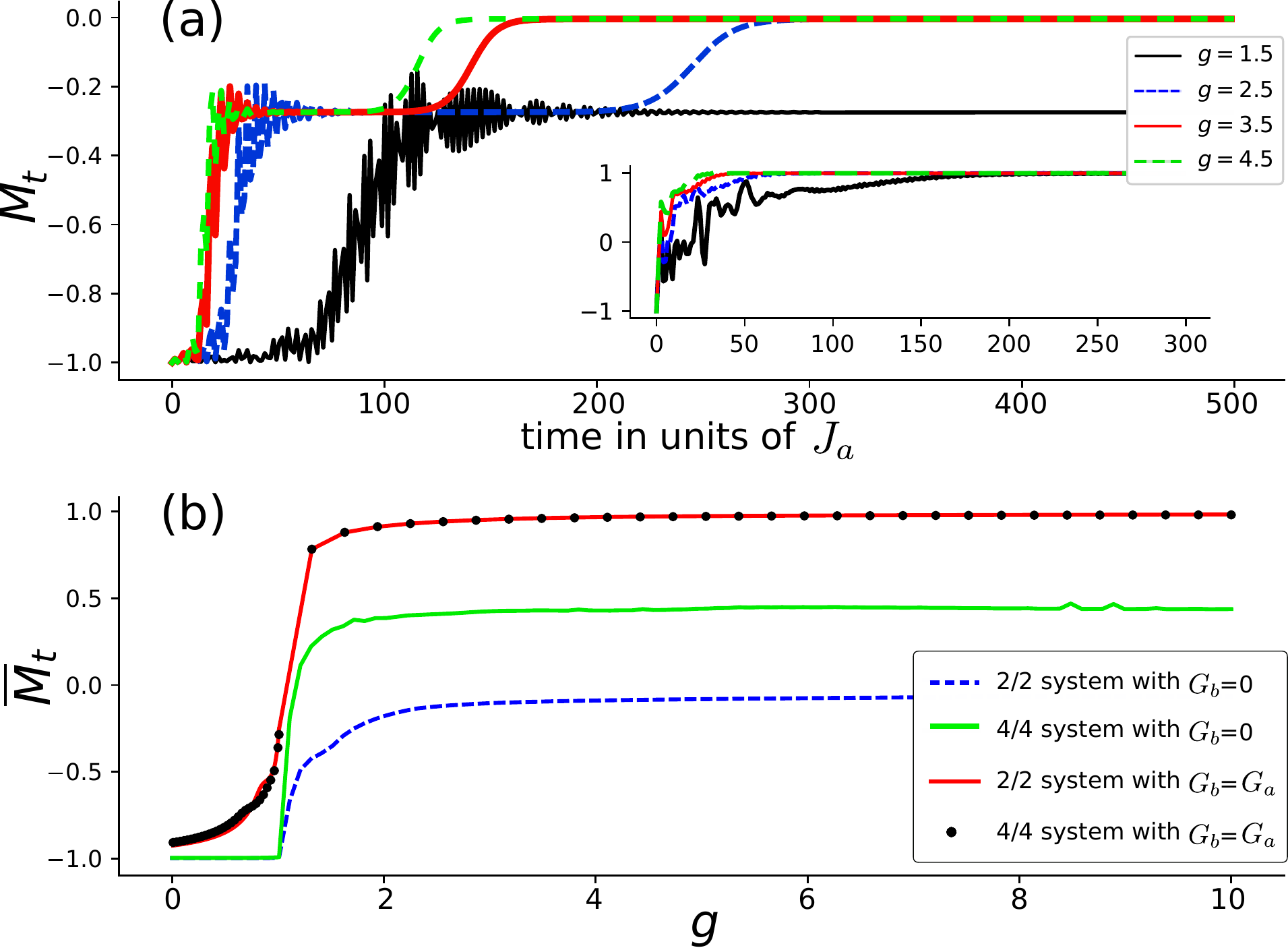}
\caption{ \label{fig:3}(Color Online) (a) The main panel shows the time evolution
  of the population inversion for the $2/2$ system, with $\Gamma_b = 0.8\times 10^{-3}$ and $\Gamma_a = g \times 0.8\times 10^{-3}$. For values 
of $g>0$ the initially locked state $|11;00\rangle$ evolves toward the 
Bell-like state $|\psi_{\rm Bell}\rangle = \frac{1}{\sqrt{2}}(|10;01\rangle + |01;10\rangle)$. The larger the value of 
$g$ the faster we reach $|\psi_{\rm Bell}\rangle$. The inset shows the 
results taking the $G_x$-terms into account with $G_b=G_a=0.09$. We see that 
the saturation is no longer about $\overline M_t \rightarrow 0$ but $\overline 
M_t \rightarrow 1$, which implies that the asymptotic state is no longer 
of the Bell type but a Mott state $\ket{\psi_{\rm Mott}}= \ket{00;11}$. 
(b) Longtime average of $M_t$ as a function of $g$. The BH parameters are the 
same as in Fig. \ref{fig:2}. For the $2/2$ system 
(dashed line), we 
reach a saturation point at $\overline{M}_t \approx 0$. However, for $4/4$, the system 
(green line) reaches the saturation point at $\overline{M}_t \approx 0.5$, 
and the final state is a superposition dominated by the state 
$|1000;2010\rangle$. When $G_b=G_a$, $\overline{M}_t \approx 1$ in the $2/2$ 
system (red line) and in the $4/4$ system (circles).
By inspection we find that the asymptotic state takes the form of a upper-band 
Mott state in these cases.
}
\end{figure}

To start, we evolve a lower-band Mott state for different ratios of the dissipation rates of every
band $g\equiv\Gamma_a/\Gamma_b$.  In Fig.~\ref{fig:3}-(a) main panel, we present the results for the minimal
system $N/L=2/2$. We see that for values of $g>1$, $M_t$ approaches zero in two steps. Firstly, it
goes to $M_t \approx -0.2$ oscillating rapidly, and then after a time interval $M_t$ reaches (smoothly)
the interband balance population value $M_t\approx 0$. At this point, the resulting asymptotic 
state takes the form $|\psi_{t\gg 1/|J_a|}\rangle = \frac{1}{\sqrt{2}}(|10;01\rangle + e^{i\phi_t}|01;10\rangle)$, 
at which the dynamics remains frozen. This kind of state is also seen in the full
model studied in Ref.~\cite{parra16}. There it was associated with a symmetry 
related to the inversion of the on-site population of the bands. 
The structure of the $|\psi_{t\gg 1/|J_a|}\rangle$ resembles a Bell state.

Following \cite{parra16}, this association becomes clearer when transforming the  
Fock basis to a new picture therein called the $w_l$-representation. It follows
that we can construct new states that have the form $|w_1\cdots w_L\rangle$, with
$w_l= n^b_l-n^a_l \in [-N,N]$. Therefore, the asymptotic state can be uniquely written 
as 
\begin{eqnarray} \label{eqn:04}
 |\psi_{t\gg 1/|J_a|}\rangle &=& \frac{1}{\sqrt{2}}(|10;01\rangle + e^{-i \phi_t}|01;10\rangle)\notag\\
 &=&\frac{1}{\sqrt{2}}(|+1;-1\rangle + e^{-i \phi_t} |-1;+1\rangle) \,,
\end{eqnarray}
with the phase factor obtained from the numerics. This notation is only helpfull if the 
states involved in the dynamics fulfill the condition $n^a_l+n^b_l=1$. The Bell-like structure 
of the state clearly appears and we can ensure that the asymptotic state is maximally 
entangled. The inclusion of the $W_x$-term does not change the shape of the asymptotic state
but the transient dynamics, i.e., the stabilization time as the previous section.

In the case $G_b=G_a\neq 0$, that is, when the interband coupling from the upper to lower band
  is similarly constrained as for the reverse process, the asymptotic state is 
given by a Mott-like state in the upper band,
  i.e., $\ket{\psi_{t\gg 1/|J_a|}}\sim \ket{000\cdots;111\cdots}$.
  Note that, since this also happens under the condition $\Gamma_a>\Gamma_b$, the upper band Mott state is the most stable
  state possible in the asymptotic regime. Here, no $G_x$-coupling will produce a transition between the
  bands because at least one doublon in the upper band has to be dynamically created. However, higher single
  site occupation will be removed rapidly by the dissipation. In the inset of 
Fig.~\ref{fig:3}-(b) we show the saturation of the
  population inversion in time, which goes to $\overline{M}_t\rightarrow 1$ for the case $G_a=G_b$. After numerical inspection
  of the expansion coefficients,  in both cases, i.e. $N/L=\{2/2,4/4\}$, the asymptotic state
  is $\ket{\psi_{t\gg 1/|J_a|}}\approx \ket{000\cdots;111\cdots}$.
   
  In Fig. \ref{fig:3}-(b) we plot the average value of $\overline{M}_t$ for
  $N/L=2/2$ (black line) and $N/L=4/4$ (green line) with and without the $G_b$-coupling
  term. The Mott state is still locked by the Zeno effect when $g\in [0,1]$ while for
  $g>1$ the interband dynamics is no longer prohibit, i.e., Zeno dynamics is no longer seen.
  For $N/L=4/4$, the saturation values for $\overline{M}_t$ is far from the perfect interband
  balance $\overline{M}_t=0$ and the final superposition state is
  mainly dominated by the state $|1000;2010\rangle$. Despite that there is still entanglement in the system,
  it is difficult to see a relevant structure of the asymptotic state in 
contrast to what is found in Ref.~\cite{parra16}.
  However, if $G_b\neq 0$ the upper-band Mott state continues to show up 
for $N/L=4/4$. 

\section{Conclusions}
\label{sec:conclusions}

We presented an effective Hamiltonian model for a two-band BH system in the presence 
of correlated interband tunneling processes. Our model accounts for the regime
 in which the doublon formation in the bottom band induces interband transport.
Motivated by some recent results on quantum many-body systems \cite{kordas}, we  
opened the system, to explore the interband dynamics. With this model, 
we built an scheme for the preparation of superposition states. 
Additionally, we found that for the smallest non-trivial system our scheme allows us to drive the 
system towards a state that resembles a Bell state in the case of strongly constrained
interband coupling from the lower to the upper band, while if both direction in the interband
coupling are equally constrained, an upper-band Mott state occurs in the asymptotic dynamics.
We presented signals of frozen quantum dynamics in a partially opened system, that is, top band open, 
bottom band closed, and we were able to relate that with the quantum Zeno 
effect in interband transport, see also \cite{zenopisa}. The interband dynamics can be unfrozen by the controlled opening also of the bottom band. 
Finally, we have shown that the quantum Zeno effect in the interband dynamics 
still appears for larger systems. 

\section{Acknowledgments}
The authors acknowledge financial support of the University del Valle (project CI 7996).
C. A. Parra-Murillo gratefully acknowledges financial support of COLCIENCIAS 
(grant 656), J.M. support from the Colombian Science, Technology and Inovation fund (Fondo 
CTEI-Sistema General de Regalas, contract BPIN 20113000100007) and COLCIENCIAS contract No. 71003,  and
S.W. support by the FIL2014 program of Parma University.

\end{document}